\documentclass[12pt]{article}
\usepackage[margin=1in]{geometry}

\usepackage{amssymb}
\usepackage{amsmath}
\usepackage{amstext}
\usepackage{graphicx,epsfig}
\usepackage{epsfig}
\usepackage{verbatim} 
\usepackage{fancyhdr}
\usepackage{fancybox}
\usepackage{color}
\usepackage{ulem,bbold}
\usepackage{enumitem}
\usepackage{subfigure}
\usepackage{bbm}
\usepackage{parskip}
\usepackage{cite}

\newcommand{\Comment}[1]{{}}
\definecolor{MyDarkBlue}{rgb}{0.15,0.15,0.45}
\usepackage[linktocpage=true]{hyperref}
\hypersetup{
colorlinks=true,
citecolor=MyDarkBlue,
linkcolor=MyDarkBlue,
urlcolor=MyDarkBlue,
}

\newcommand{\be}{\begin{equation}}
\newcommand{\ee}{\end{equation}}
\newcommand{\bea}{\begin{eqnarray}}
\newcommand{\eea}{\end{eqnarray}}
 
\newcommand{\e}{{ \rm{e}}}

\numberwithin{equation}{section}

\begin{document}

\thispagestyle{empty}

\begin{center}
{\Large \bf{Non-Singular Black Holes in Massive Gravity:}}
{\Large \bf{Time-Dependent Solutions}}
\end{center} 

 \vspace{1truecm}
 
\centerline{{\large  {Rachel A. Rosen}}}

\centerline{{ \it Department of Physics, Columbia University,}}
\centerline{{\it  New York, NY 10027, USA}}

\vspace{1cm}

\begin{abstract} 

When starting with a static, spherically-symmetric ansatz, there are two types of black hole solutions in dRGT massive gravity: (i) exact Schwarzschild solutions which exhibit no Yukawa suppression at large distances and (ii) solutions in which the dynamical metric and the reference metric are simultaneously diagonal and which inevitably exhibit coordinate-invariant singularities at the horizon.  In this work we investigate the possibility of black hole solutions which can accommodate both a nonsingular horizon and Yukawa asymptotics.  In particular, by adopting a time-dependent ansatz, we derive perturbative analytic solutions which possess non-singular horizons.  These black hole solutions are indistinguishable from Schwarzschild black holes in the limit of zero graviton mass.   At finite graviton mass, they depend explicitly on time.  However, we demonstrate that the location of the apparent horizon is not necessarily time-dependent, indicating that these black holes are not necessarily accreting or evaporating (classically).  In deriving these results, we also review and extend known results about static black hole solutions in massive gravity.

\end{abstract}

\newpage

\thispagestyle{empty}
\tableofcontents
\newpage
\setcounter{page}{1}
\setcounter{footnote}{0}

\section{Introduction and Summary}
Viable theories of gravity almost inevitably require a spin-2 particle at their foundation.  It remains an open question whether or not this spin-2 particle is strictly massless as in General Relativity (GR) or if it can be massive.  Until recently, it was unknown if fundamental, Lorentz-invariant massive spin-2 particles were even theoretically viable.  In 2010, de Rham, Gabadadze and Tolley (dRGT) \cite{deRham:2010ik,deRham:2010kj} succeeded in constructing a low energy, Lorentz-invariant theory of a massive spin-2 field that was free of the pathology known as the Boulware-Deser ghost \cite{Boulware:1973my} (see, \cite{Hassan:2011hr}).  Since then there has been considerable effort to understand the implications of this theory. (For reviews, see, \cite{Hinterbichler:2011tt,deRham:2014zqa}.)  In particular, black hole solutions provide an important test of both the theoretical and phenomenological viability of dRGT massive gravity.

Static black hole solutions in dRGT massive gravity have been studied extensively in the literature, with many central results given in  \cite{Koyama:2011xz,Nieuwenhuizen:2011sq,Koyama:2011yg,Gruzinov:2011mm,Comelli:2011wq,Berezhiani:2011mt,Volkov:2012wp,Sbisa:2012zk,Gratia:2012wt,Chiang:2012vh,Mirbabayi:2013sva,Volkov:2013roa,Tasinato:2013rza,Babichev:2013una,Brito:2013xaa,Arraut:2013bqa,Kodama:2013rea,Renaux-Petel:2014pja,Babichev:2014oua,Volkov:2014ooa,Babichev:2015xha}.  Let us summarize some of the main findings and refer to \cite{Babichev:2015xha} for a recent review.  Starting with a static, spherically symmetric ansatz for the dynamical metric and taking the reference metric to be Minkowski, dRGT massive gravity admits two branches of solutions:
\begin{itemize}
\item  On the first branch, the dynamical and reference metrics are not simultaneously diagonal and exact solutions of the Schwarzschild or A/dS-Schwarzschild type can be found.  These solutions exhibit no Yukawa-type suppression at large distances and, indeed, on this branch the graviton mass can be shown to vanish around flat backgrounds.  This indicates that this branch of solutions is infinitely strongly coupled and is not smoothly connected to the usual physical massive gravity theory.  

\item On the second branch, the dynamical metric is simultaneously diagonal with the reference metric.  This branch is also problematic since it has been shown that, in theories with two static, bi-diagonal metrics, if a Killing vector $\partial_t$ is null at some radius $r=r_H$ with respect to one metric, then it must also be null at $r=r_H$ with respect to the second metric in order to avoid coordinate-invariant singularities at the horizon \cite{Deffayet:2011rh}.  Since we take the reference metric of dRGT massive gravity to be strictly Minkowski with no such horizon, the solutions on this branch inevitably contain singularities at the horizon of the dynamical metric.
\end{itemize}

In this work, we investigate the possibility of new black hole solutions with both (i) non-zero Fierz-Pauli mass \cite{Fierz:1939ix} around flat spacetime and (ii) non-singular horizons.  The first criteria is significant because it puts us in the relevant region of parameter space.  I.e., we restrict our attention to theories in which the massive graviton propagates the appropriate five degrees of freedom around flat spacetime.  The second criteria is significant because we wish to understand whether or not black holes in the massless limit of massive gravity look arbitrarily close to black holes in General Relativity or if there is necessarily a discontinuity.  The possibility of a discontinuity between the massless limit of massive gravity and GR was first pointed out by van Dam, Veltman and Zakharov (vDVZ) \cite{vanDam:1970vg,Zakharov:1970cc} in the context of the linear theories.  However, it was pointed out by Vainshtein \cite{Vainshtein:1972sx} that the non-linearities in massive gravity could act to recover the predictions of General Relativity at short distances.  Indeed, dRGT massive gravity has such a mechanism for a wide range of parameter space when considering general astrophysical sources, i.e., objects without a horizon.  However, the presence of coordinate-invariant singularities at the black hole horizon for arbitrarily small graviton mass would indicate the reappearance of a discontinuity for black hole solutions.  In principle, this could make massive gravity phenomenologically distinguishable from General Relativity, even at arbitrarily small graviton mass.  This issue is particularly timely, given upcoming experimental results from, e.g., LIGO or the Event Horizon Telescope, which could potentially differentiate between these scenarios. 

To go beyond the two known branches of solutions and search for black holes which satisfy the criteria given above, we relax the assumption of a static ansatz and look instead for time-dependent black hole solutions.  That such an assumption could resolve the singularities was first suggested in \cite{Mirbabayi:2013sva}.  The time dependence is introduced by assuming that, in the limit of zero graviton mass, the dynamical metric is exactly Schwarzschild while the reference metric is a time-dependent coordinate transformation of Minkowski.  With this ansatz, the equations of motion can be satisfied (away from the so-called ``minimal model").  The equations of motion determine the form of the coordinate transformation.  We solve these equations perturbatively to find coordinates very similar to Kruskal-Szekeres coordinates.  In the limit of zero graviton mass, this black hole is identical to that of GR.  At small finite mass, the black hole solutions become time-dependent.  However, we will show that the location of the apparent horizon is not time-dependent at leading order in small graviton mass, indicating that these black holes are not necessarily accreting or evaporating (classically).

{\it This paper is organized as follows:}  In section \ref{background} we briefly introduce dRGT massive gravity.  In section \ref{staticsol} we review and expand on black hole solutions in massive gravity for a static, spherically symmetric ansatz.  In particular, in section \ref{branchI} we review the branch of solutions that yields exact Schwarzschild black holes and demonstrate the vanishing of the Fierz-Pauli mass.   In section \ref{branchII} we review the bi-diagonal branch of solutions.  We focus on the massless limit in order to understand the appearance or failure of the Vainshtein mechanism.  We treat the minimal, next-to-minimal and non-minimal models separately.  For the minimal model, we find an analytic solution in the massless limit that explicity demonstrates the absence of a Vainshtein mechanism, as expected.  For the next-to-minimal and non-minimal models, Schwarzschild solutions can be found in the strictly massless limit, naively indicating a functioning Vainshtein mechanism.  However, these solutions contain coordinate invariant singularities at the horizon for arbitrarily small graviton mass.  This indicates that the massless limit is discontinuous and that these black holes can be distinguished from GR black holes for arbitrarily small mass.

In section \ref{timedep} we again consider the next-to-minimal model.  Here, however, we relax the assumption of a static ansatz and we allow the reference metric to be a time-dependent coordinate transformation of Minkowski.   We solve the equations of motion perturbatively and demonstrate that, to at least 5th order in our expansion, these solutions contain no singularities at the horizon.  We derive the finite mass corrections to the dynamical metric and show that the black hole solution becomes explicitly time-dependent.  We then show that the location of the apparent horizon is independent of time, to leading order in small mass.


\section{Background}
\label{background}
Our starting point is the dRGT Lagrangian for ghost-free massive gravity \cite{deRham:2010kj}:
\be
\label{L}
{\cal L} = \frac{M_{Pl}^2}{2}\sqrt{-g}\left[R-2 m^2\sum_{n=0}^4 \beta_n S_n(\sqrt{g^{-1} f}) \right] \, .
\ee
This Lagrangian contains the usual Einstein-Hilbert kinetic term for the dynamical metric $g_{\mu\nu}$. In addition, there is a potential term containing no derivatives of the dynamical metric but which depends explicitly on a non-dynamical reference metric $f_{\mu\nu}$.  The reference metric breaks the diffeomorphism invariance of the $m = 0$ theory.  However, if we take the reference metric to be Minkowski $f_{\mu\nu}=\eta_{\mu\nu}$ then the theory is Lorentz invariant.  In this paper we will consider this case.

In the potential term, the $S_n$ are the $n$-th elementary symmetric polynomials of the eigenvalues of the matrix square root of $g^{\mu\lambda}f_{\lambda\nu}$.  They are given by
\bea
\label{potential}
\begin{array}{l}
S_0 (\mathbb{X})= 1  \, ,  \vspace{.1cm} \\
S_1(\mathbb{X})= [\mathbb{X}]  \, ,  \vspace{.1cm} \\
S_2(\mathbb{X})= \tfrac{1}{2}([\mathbb{X}]^2-[\mathbb{X}^2]) \, ,  \vspace{.1cm} \\
S_3(\mathbb{X})= \tfrac{1}{6}([\mathbb{X}]^3-3[\mathbb{X}][\mathbb{X}^2]+2[\mathbb{X}^3]) \, ,  \vspace{.1cm} \\
S_4(\mathbb{X})=\tfrac{1}{24}([\mathbb{X}]^4-6[\mathbb{X}]^2[\mathbb{X}^2]+3[\mathbb{X}^2]^2   
+8[\mathbb{X}][\mathbb{X}^3]-6[\mathbb{X}^4])\, .
\end{array}
\eea
The square brackets denote the trace of the enclosed matrix.  

The $\beta_n$ are free coefficients.  If we expand the dynamical metric around flat spacetime $g_{\mu\nu} = \eta_{\mu\nu} + 2 h_{\mu\nu}/M_{Pl}$ then the requirement of no tadpoles gives a condition on the $\beta_n$:
\be
\label{cc}
\beta_0 + 3 \beta_1 + 3 \beta_2 + \beta_3 = 0 \, .
\ee
In other words, this condition sets to zero the cosmological constant term coming from the potential.  Assuming this condition, the correct normalization of the mass $m^2$ means that
\be
\label{mass}
\beta_1 +2\beta_2 +\beta_3 =1 \, .
\ee
I.e., this condition guarantees that around flat space, at lowest order in the fields, the Lagrangian \eqref{L} reduces to the linear Fierz-Pauli Lagrangian for the free massive graviton, with mass term 
\be
\label{FP}
 \frac{m^2}{2}(h_{\mu\nu}h^{\mu\nu}-h^\mu_{\, \mu}h^\nu_{\, \nu}) \, .
\ee
In addition, $\beta_4$ multiplies a non-dynamical term and can be set to zero, leaving two free parameters among the $\beta_n$.  In what follows, we will often take equations \eqref{cc} and \eqref{mass} to be the defining equations for $\beta_0$ and $\beta_1$ and we will take $\beta_2$ and $\beta_3$ to be free parameters.

\section{Static Solutions}
\label{staticsol}
\subsection{Branches of Solutions}
Let us start by considering the linear Fierz-Pauli theory of massive gravity with mass given by \eqref{FP} in the presence of a point source of mass $M$.  We adopt spherical coordinates $(\tau,\rho,\theta,\phi)$ and a static, spherically symmetric ansatz for the dynamical metric $g_{\mu\nu}$:
\bea
\label{ansatz}
\begin{array}{l}
ds_g^2 = - A_{00}^2(\rho) d\tau^2+2A_{01}(\rho)d\tau d\rho+A_{11}^2(\rho) d\rho^2+\rho^2 A_{22}^2(\rho)\, d\Omega^2  \, , \vspace{0.3cm}\\
ds_f^2 = - d\tau^2+d\rho^2+ \rho^2\, d\Omega^2  \, ,
\end{array}
\eea
where $d\Omega^2$ is the metric of the unit 2-sphere, $d\Omega^2 = d\theta^2 +\sin^2 \theta\, d\phi^2$.  The solutions to the equations of motion are given by (see, e.g., \cite{Hinterbichler:2011tt})
\bea
\label{Yukawa}
\begin{array}{l}
 A_{00}^2(\rho)  =  1-\dfrac{8GM}{3} \dfrac{e^{-m\rho}}{\rho}\, , \vspace{0.3cm}\\
 A_{01}(\rho) =  0\, , \vspace{0.3cm}\\
 A_{11}^2(\rho)  = 1-\dfrac{8GM}{3} \dfrac{e^{-m\rho}}{\rho} \dfrac{1+m\rho}{m^2\rho^2} \, , \vspace{0.3cm}\\
 A_{22}^2(\rho) = 1+\dfrac{4GM}{3} \dfrac{e^{-m\rho}}{\rho} \dfrac{1+m\rho+m^2\rho^2}{m^2\rho^2}  \, . \\
\end{array}
\eea
As expected, the existence of the mass term at the linear level gives rise to Yukawa suppression of the potential at large $\rho$.

We now wish to find vacuum solutions to the full non-linear theory \eqref{L}.  Our expectation is that these should agree with the linear solutions \eqref{Yukawa} at large distances from the source.  We start from the same generic, static and spherically symmetric ansatz \eqref{ansatz}.  The nonlinear equations of motion take the form:
\be
\label{EOM}
G_{\mu\nu}+m^2 \, {\cal T}_{\mu\nu} = 0 \, .
\ee
Here ${\cal T}_{\mu\nu}$ represents all the contributions to the equations of motion coming from the potential term in \eqref{L}.  From the $(0,1)$ component of the equations of motion \eqref{EOM}, it is straightforward to see that there are two possible branches of solutions:
\begin{itemize}
\item {\bf BRANCH I:}
\be
\label{b1}
\beta_1 A_{22}(\rho)^2 + 2 \beta_2 A_{22}(\rho) +\beta_3 =0 \, .
\ee
On this branch exact solutions can be readily obtained (for early work, see, \cite{Koyama:2011xz,Nieuwenhuizen:2011sq,Koyama:2011yg}).  However, this branch corresponds to setting the mass of quadratic fluctuations to zero around flat space.   Thus this theory is infinitely strongly coupled: i.e., it would appear to propagate only two degrees of freedom around flat space but would propagate five degrees of freedom around curved backgrounds.  Correspondingly, it does not possess the Yukawa type asymptotics \eqref{Yukawa} that we desire.

\item {\bf BRANCH II:}
\be
\label{b2}
A_{01}(\rho)=0 \, .  
\ee
On this branch of solutions the two metrics are required to be simultaneously diagonal.  As discussed in the introduction, this generically leads to coordinate-invariant singularities at the black hole horizon \cite{Deffayet:2011rh}.  We will demonstrate this explicitly below.

\end{itemize}

In what follows we will review these two branches and their features.  Then, in section \ref{timedep} we will consider alternative solutions resulting from a time-dependent ansatz.

\subsection{Branch I: Exact Schwarzschild Solutions}
\label{branchI}
We consider first the branch defined by equation \eqref{b1}:
\be
\label{b12}
\beta_1 A_{22}(\rho)^2 + 2 \beta_2 A_{22}(\rho) +\beta_3 =0 \, .
\ee
On this branch, exact solutions can be readily obtained.  For convenience, we transform from the coordinates $(\tau, \rho)$ of \eqref{ansatz} in which the reference metric is explicitly Minkowski to new coordinates $(t, r)$.  The unique solutions are given by:
\bea
\label{bh1}
\begin{array}{l}
ds_g^2 = -V(r) dt^2+ \frac{1}{V(r)}dr^2+ r^2  \, d\Omega^2   \, , \vspace{0.3cm}\\
ds_f^2 = - C_0^2\, dt^2+C_0 \sqrt{U(r)}\,dt dr+\Large(C_1^2 - U(r)\Large)\,dr^2+ C_1^2\, r^2\, d\Omega^2 \, .
\end{array}
\eea
Here $C_0$ and $C_1$ are constants.  The functions $V(r)$ and $U(r)$ are given by
\be
\label{Vdef}
V(r) = 1-\frac{r_g}{r}-\frac{\Lambda}{3} r^2 \, ,
\ee
\be
U(r) = \left( \frac{C_0^2}{V(r)^2} - \frac{C_1^2}{V(r)}\right)\left(1-V(r) \right) \, .
\ee
The reference metric can be transformed back to Minkowski by the change of variables:
\be
\label{transf}
\tau(t,r) = C_0\, t - \int dr \sqrt{U(r)} \, ,~~~~ \rho(t,r) = C_1\, r \, ,
\ee
\be
\Rightarrow ds_f^2 = - d\tau^2+d\rho^2+ \rho^2\, d\Omega^2  \, .
\ee
Thus, by considering equations \eqref{bh1} and \eqref{Vdef} we see that these solutions describe {\it exact} Schwarzschild and A/dS-Schwarzschild solutions with cosmological constant $\Lambda$.  The reference metric is flat.  Notably, these solutions lack the expected Yukawa suppression at large distances \eqref{Yukawa}.

To understand this, we note that the condition that defines this branch \eqref{b12}, translates into the following relation
\be
\label{Crho}
\beta_1 + 2 \,C_1 \, \beta_2 +C_1^2\, \beta_3 =0 \, ,
\ee
while $\Lambda$ is defined by the relation
\be
m^2 (\beta_0 +3\, C_1\, \beta_1+3\, C_1^2\, \beta_2+ C_1^3\, \beta_3) = \Lambda \, .
\ee
we see that $C_1$ can be absorbed into the definition of the $\beta_n$.  Thus, the condition \eqref{Crho} (or, equivalently \eqref{b12}) appears to correspond to setting the quadratic Fierz-Pauli mass to zero, as was first pointed out in \cite{Comelli:2011wq}.  

Given the presence of $C_0$, let us examine this statement closely.  Let us take $\Lambda = 0$ and consider the regime in which the dynamical metric is approximately flat $g_{\mu\nu} \simeq \eta_{\mu\nu}$, i.e., when $r_g \ll r $.  In this regime the reference metric takes the following form:
\be
\label{fscale}
ds_f^2 = - C_0^2\, dt^2+C_1^2 \,dr^2+ C_1^2\, r^2\, d\Omega^2 \, .
\ee
Let us revisit the mass normalization condition \eqref{mass} for this reference metric.  We consider two cases:
\begin{itemize}
\item  {\bf CASE 1:}  $C_0 = C_1$.  In this case the reference metric is given by $f_{\mu\nu} = C_1\, \eta_{\mu\nu}$.  Expanding the dynamical metic $g_{\mu\nu}$ around $\eta_{\mu\nu}$, the condition for no tadpoles becomes
\be
\label{cc2}
\beta_0 + 3 \, C_1 \beta_1 + 3\,  C_1^2 \beta_2 +C_1^3 \beta_3 =0 \, , 
\ee
while the mass normalization condition is given by
\be
\label{m2}
\beta_1 +2\, C_1\beta_2 +C_1^2\beta_3 =1 \, .
\ee
If we compare equation \eqref{m2} with the defining condition for this branch of solutions \eqref{Crho}, we see that the solutions \eqref{bh1} correspond to having a zero Fierz-Pauli mass for the metric fluctuation around flat space.

\item {\bf CASE 2:}  $C_0 \neq C_1$.  If we now expand the dynamical metric $g_{\mu\nu}$ around $\eta_{\mu\nu}$  we find the condition of no tadpoles requires both
\be
\label{cc3}
\beta_0 + 3\, C_1 \beta_1 + 3\, C_1^2 \beta_2 +C_1^3 \beta_3 =0 \, , ~~{\rm and}~~ \beta_1 +2\, C_1\beta_2 +C_1^2\beta_3=0 \, .
\ee
Assuming these conditions, the Fierz-Pauli structure of the mass is altered.  At quadratic level one finds a mass term for the spatial components of the metric alone:
\be
\label{cc4}
\sim m^2  C_1 (C_1-C_0)(\beta_2+C_1 \beta_3)(h_{ij}h^{ij}-h^i_{\ i}h^j_{\ j}) \, .
\ee
Thus when $C_0 \neq C_1$ the solutions \eqref{bh1} correspond to having a zero mass for the $h_{00}$ component of the metric fluctuation.  
\end{itemize}

Given these results, it makes sense that these solutions do not have the usual Yukawa suppression at large distances.


Finally, we note that on these solutions all scalar quantities are finite:
\be
\sqrt{g^{-1}f}^{\,\mu}_{~~\mu} = C_0+3 C_1\, , ~~~~
m^2\, {\cal T}^\mu_{~~\mu} = \Lambda\, \delta^\mu_{~~\mu}  \, ,
\ee
where ${\cal T}^\mu_{~~\nu}$ is defined as in \eqref{EOM}.  In other words, no coordinate-invariant singularities exist at the horizon of these black holes.  However, while these solutions are interesting in their own right, they require a choice of parameters which results in a theory that does not propagate a usual Fierz-Pauli massive graviton around flat space.  On this branch of solutions, at higher order in perturbations or, equivalently, around curved backgrounds, the massive graviton would appear to propagate a different number of degrees of freedom than around flat space, indicating that this theory is infinitely strongly coupled.

\subsection{Branch II: Bidiagonal Solutions in the $m \rightarrow 0$ Limit}
\label{branchII}
We next consider the bidiagonal branch of solutions, defined by $A_{01}(\rho)=0$:
\bea
\label{diag}
\begin{array}{l}
ds_g^2 = - A_{00}^2(\rho) d\tau^2+A_{11}^2(\rho) d\rho^2+ \rho^2 A_{22}^2(\rho) \, d\Omega^2   \, , \vspace{0.3cm}\\
ds_f^2 = - d\tau^2+d\rho^2+ \rho^2\, d\Omega^2  \, .
\end{array}
\eea
The equations of motion simplify greatly in Schwarzschild-type coordinates \cite{Koyama:2011yg}.  We introduce a new radial coordinate $r(\rho) \equiv \rho A_{22}(\rho)$ and perform a coordinate transformation so that our ansatz \eqref{diag} becomes
\bea
\label{schwarz}
\begin{array}{l}
ds_g^2 = - B_0^2(r) dt^2+B_1^2(r) dr^2+ r^2\, d\Omega^2   \, , \vspace{0.3cm}\\
ds_f^2 = - dt^2+\rho'(r)^2 dr^2+ \rho(r)^2\, d\Omega^2 \, ,
\end{array}
\eea
where 
\bea
\begin{array}{l}
B_0(r) \equiv A_{00}[\rho(r)] \, , \vspace{0.3cm}\\
B_1(r) \equiv  \rho'(r)\, A_{11}[\rho(r)] \, .
\end{array}
\eea
Here primes denote derivatives with respect to $r$.  The anticipated asymptotic solutions \eqref{Yukawa} become
\bea
\label{Yukawa_Schwarz}
\begin{array}{l}
B_0^2(r)  \rightarrow  1-\dfrac{8GM}{3} \dfrac{e^{-mr}}{r}\, , \vspace{0.3cm}\\
B_1^2(r)  \rightarrow 1+\dfrac{4GM}{3} \dfrac{e^{-mr}}{r} (1+mr) \, , \vspace{0.3cm}\\
 \rho(r) \rightarrow r \left(1-\dfrac{2GM}{3} \dfrac{e^{-mr}}{r} \dfrac{1+mr+m^2r^2}{m^2r^2}  \right)\, . \\
\end{array}
\eea
The equations of motion \eqref{EOM} give three independent equations for the three unknown functions $B_0(r)$, $B_1(r)$ and $\rho(r)$:
\bea
\label{EOM2}
\begin{array}{rcl}
2r B'_1(r)+\big(1-m^2(\beta_0r^2+2\beta_1r \rho(r)+\beta_2\rho(r)^2)\big)B_1(r)^3&& \\
-m^2\big(\beta_1 r^2+2\beta_2r\rho(r)+\beta_3 \rho(r)^2\big)\rho'(r)B_1(r)^2-B_1(r) & = & 0 \, , \vspace{0.3cm} \\
2r B'_0(r)+\big(1-\big(1-m^2(\beta_0r^2+2\beta_1r \rho(r)+\beta_2\rho(r)^2)\big)B_1(r)^3\big)B_0(r) && \\
+m^2\big(\beta_1 r^2+2\beta_2r\rho(r)+\beta_3 \rho(r)^2\big)B_1(r)^2 & = & 0 \, , \vspace{0.3cm} \\
\big(\beta_1 r^2+2\beta_2r\rho(r)+\beta_3 \rho(r)^2\big)B'_0(r) &&\\
+2\big(\beta_1 r B_0(r)+\beta_2 (r+B_0(r)\rho(r)) +\beta_3\rho(r)\big)(1-B_1(r)) & = & 0\, .
\end{array}
\eea
The first two equations are simply the equations of motion of General Relativity plus $m^2$ corrections due to the potential \eqref{potential}.  The third equation is due to the covariant conservation of the equations of motion on-shell, i.e., $\nabla_\mu {\cal T}^\mu_{~~\nu} = 0$.  This equation is not present in General Relativity.  It persists in the $m\rightarrow 0 $ limit of massive gravity.

\subsubsection{Minimal Model: $\beta_2 = 0$, $\beta_3 = 0$}
We attempt to solve the equations of motion \eqref{EOM2} first in the simplest possible case, the so-called ``minimal model"  \cite{Hassan:2011vm}.  In this model, the helicity-0 mode of the massive graviton has no interactions in the decoupling limit and no Vainshtein mechanism is expected.  It corresponds to the choice of parameters $c_3 = 1/6$, $d_5 = -1/48$ of \cite{deRham:2010ik,deRham:2010kj} or, equivalently,
\be
\beta_0 + 3 \beta_1 + 3 \beta_2 + \beta_3 = 0 \, , ~~~~ \beta_1 + 2 \beta_2 + \beta_3 = 1 \, , ~~~~ \beta_2 = \beta_3 = 0 \, .
\ee
With this choice the equations of motion greatly simplify.  With some rearranging of \eqref{EOM2}, one finds:
\bea
\label{EOMmin}
\begin{array}{l}
8 r^2 \dfrac{B''_0(r)}{B_0(r)}+3r^3 \dfrac{B'_0(r)^3}{B_0(r)^3}+2r^2\dfrac{B'_0(r)^2}{B_0(r)^2}
+16 r \dfrac{B'_0(r)}{B_0(r)} -m^2r^2\left(1-\dfrac{1}{B_0(r)}\right)\left(\dfrac{r B'_0(r)}{B_0(r)}+2 \right)^3 = 0  \, , \vspace{0.3cm} \\
B_1(r)  =  1+\dfrac{r B'_0(r)}{2 B_0(r)} \, , \vspace{0.3cm} \\
\rho(r) = \dfrac{1}{m^2} \dfrac{B'_0(r)\left(r B'_0(r)-4B_0(r)\right)}{2\left(r B'_0(r)+2B_0(r)\right)^2} \
+\dfrac{1}{2}r \left(3-\dfrac{1}{B_0(r)}\right)  \, .
\end{array}
\eea
Once the first equation is solved for $B_0(r)$, the second two equations give $B_1(r)$ and $\rho(r)$ respectively.  The above equations are exact to all orders.  

We solve the first equation in the $m\rightarrow 0$ limit.  A peculiar feature of the minimal model is that the third equation in \eqref{EOM2} becomes independent of $\rho(r)$ when $\beta_2 = \beta_3 =0$.  Thus this equation can be used to solve for $B_1(r)$ in term of $B_0(r)$ and $B_0'(r)$.  But this equation is absent in General Relativity as it arises from the constraint $\nabla_\mu {\cal  T}^\mu_{~~\nu} = 0$.  Thus the resulting solutions will be different from GR, even in the massless limit.  In other words, the fact that the minimal model solutions have no Vainshtein mechanism can be read off from the equations of motion, as was observed in \cite{Renaux-Petel:2014pja}.

Setting $m=0$ in the first equation of \eqref{EOMmin} and solving, we find:\footnote{We are grateful to Riccardo Penco for suggestions in deriving this solution.}
\be
\label{minsol2}
B_0(r) = \frac{K_1}{r^{4/3}}\, \left(\frac{3}{1+2\cos\left[\tfrac{1}{3}\arctan\left(\frac{-2\sqrt{\frac{K_2}{r}\left(1- \frac{K_2}{r}\right)}}{1-2\frac{K_2}{r} }\right)\right]}-1 \right)^{\!\!-4/3}+{\cal O}(m^2) \, .
\ee
where $K_1$ and $K_2$ are constants of integration.  The solution is real for $0<\frac{K_2}{r} <1$.  Because this solution exhibits no Vainshtein mechanism, linearizing first and then taking the $m\rightarrow 0$ limit gives the same result as first taking $m\rightarrow 0$ and then linearizing.  Because of this, we can match the large distance behavior of the solution \eqref{minsol2} with Yukawa asymptotics  \eqref{Yukawa_Schwarz}.  This fixes the two integration constants
\be
K_1= \left(\frac{GM}{3} \right)^{4/3}\, , ~~~~K_2 = \frac{9}{4}\, G M \, .
\ee
This solution can now be substituted into the remaining equations of motion \eqref{EOMmin} to find $B_1(r)$ and $\rho(r)$.

In Figure 1 we compare $-g_{00}(r) =  B_0(r)^2$ with the Schwarzschild solution $-g_{00}(r) = 1-\frac{2 G M}{r} $.  The minimal massive gravity model is plotted in dark gray while the Schwarzschild solution is in light gray.  In the right hand plot we zoom in, close to $r = 2 GM$.  Both solutions asymptote to 1 for large $r$.  The observable difference between the two solutions at large $r$ is a direct manifestation of the usual $\frac{4}{3}$ factor of the vDVZ discontinuity.  At short $r$, the discrepancy is due to the absence of the Vainshtein mechanism.

\begin{figure}[t!]
\centering
\vspace{-2cm}
\includegraphics[width=8cm]{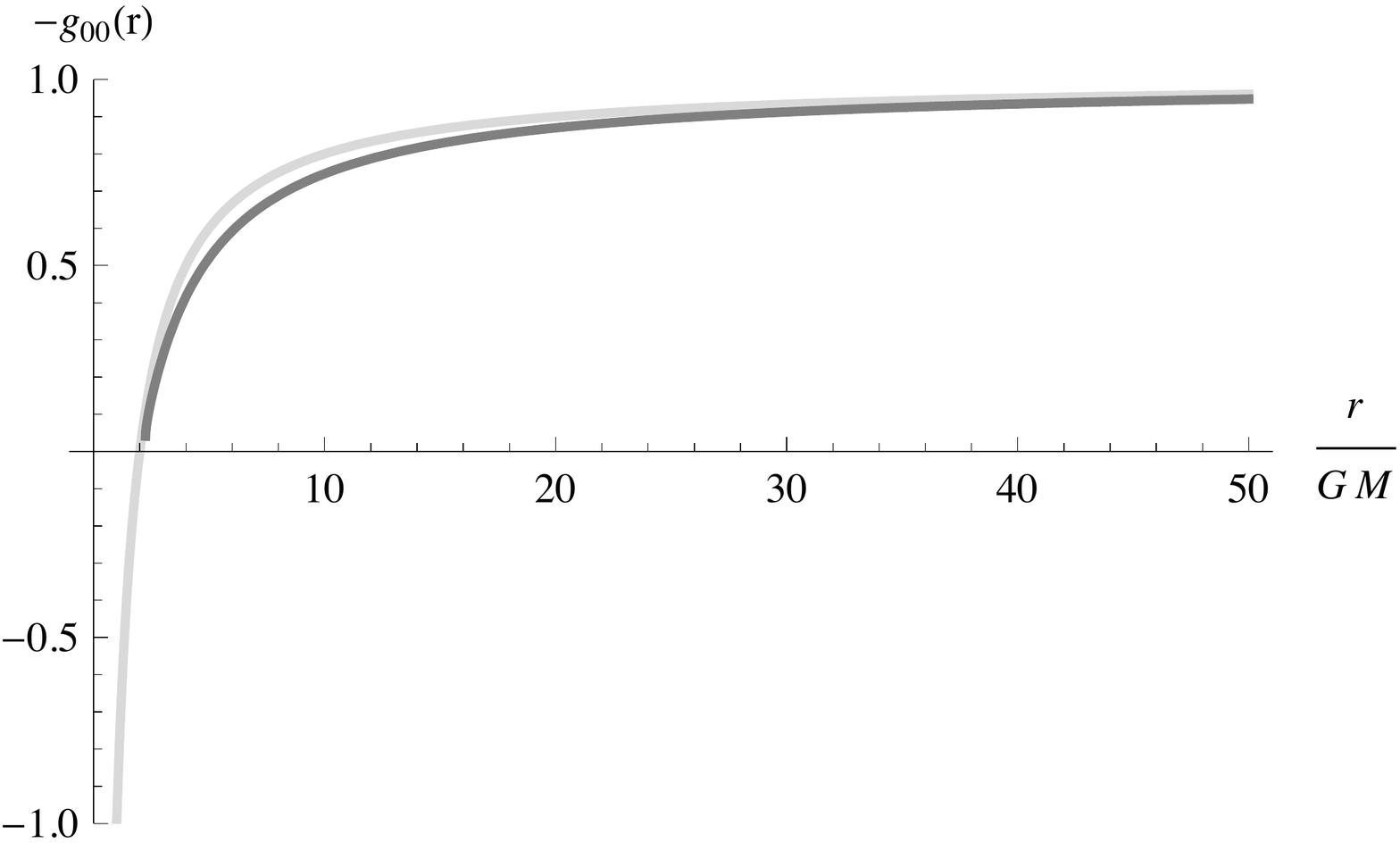} \hfill
\includegraphics[width=8cm]{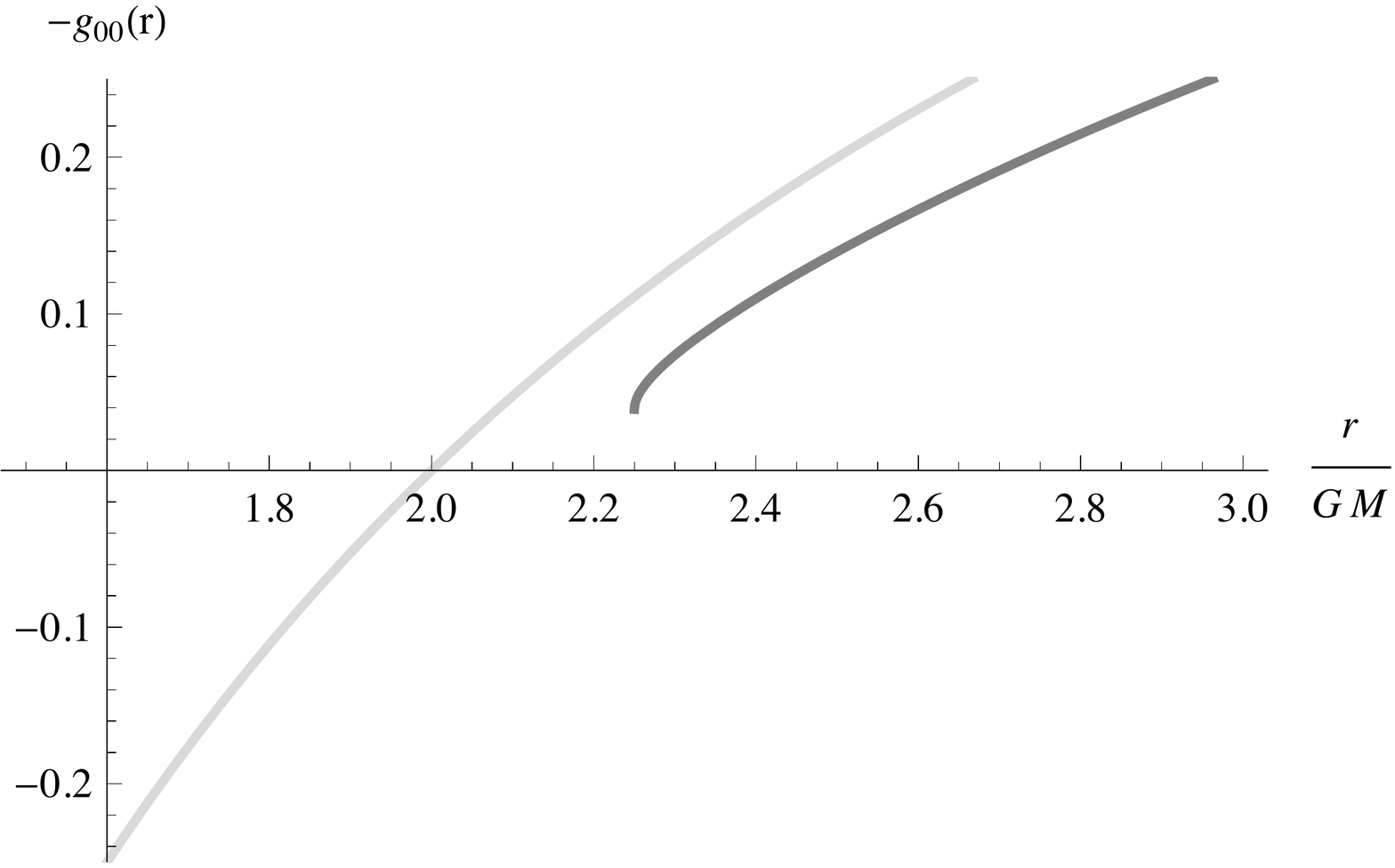} \vspace{-2cm}
\caption{Comparison of the $g_{00}(r)$ component of the metric for black holes in minimal massive gravity (dark gray) and Schwarzschild black holes (light gray).}
\end{figure}

While the $g_{00}$ component of the Schwarzschild solution passes smoothly through zero at $r = 2 G M$, for the minimal model we have abrupt behavior: at distances shorter than $r = \frac{9}{4}\, G M $, the solution becomes imaginary.  At $r =\frac{9}{4}\, G M$, we have $- g_{00}(r) = \frac{2^8}{3^8}$.  In addition, all the relevant scalars of this theory are finite at this point:
\be
R ={\cal  T}^{\mu}_{~~\mu} = 3 \sqrt{g^{-1}f}^{\,\mu}_{~~\mu} \rightarrow -\frac{32}{27} \frac{1}{(G M)^2} \, .
\ee
In other words, there are no singularities simply because this solution has no horizon.

Suggestively, the value of the radius $r =\frac{9}{4}\, G M$ is the same as that for the Buchdahl bound which gives the minimum radius of a star with finite pressure under other generic assumptions.  However, it's unclear if these solutions are physically meaningful, given that they cannot obviously be continued for $r < \frac{9}{4}\, G M$.

\subsubsection{Next-to-Minimal Model: $\beta_2 \neq 0$, $\beta_3 = 0$}
We next consider nonlinear bi-diagonal solutions in the next-to-minimal model, characterized by parameters  $\beta_2 \neq 0$, $\beta_3 = 0$.  This corresponds to  $c_3 = -8 d_5$ in the parametrization of \cite{deRham:2010ik,deRham:2010kj}.  In particular, we enforce
\be
\beta_0 + 3 \beta_1 + 3 \beta_2 + \beta_3 = 0 \, , ~~~~ \beta_1 + 2 \beta_2 + \beta_3 = 1 \, , ~~~~ \beta_3 = 0 \, ,
\ee
and parametrize all nonzero $\beta_n$ in terms of $\beta_2$.  The third equation of motion given in \eqref{EOM2} can now be solved explicitly for $\rho(r)$ in terms of $B_0(r)$ and $B_1(r)$:
\be
\rho(r) = \frac{r}{\beta_2}\frac{\left((2\beta_2-1)B_0(r)-\beta_2\right)+\frac{1}{2}(2\beta_2-1)r B_0'(r)}{B_0(r)(1-B_1(r))+r B_0'(r)} \, .
\ee
We note the lack of explicit $m$ dependence in this equation as compared to the equation for $\rho(r)$ in the minimal model \eqref{EOMmin}.  Because of this, in the next-to-minimal model, the equations of motion contain a Schwarzschild solution for $B_0(r)$ and $B_1(r)$ in the $m \rightarrow 0$ limit.  In other words, the next-to-minimal model (naively) exhibits a Vainshtein mechanism.

In particular, to leading order in small $m$, we find solutions of the form
\bea
\label{EOMnextomin}
\begin{array}{l}
B_0(r) =  \sqrt{1-\frac{r_g}{r}}+{\cal O}(m^2) \, , \vspace{0.3cm} \\
B_1(r)  = \frac{1}{\sqrt{1-\frac{r_g}{r}}} +{\cal O}(m^2)  \, , \vspace{0.3cm} \\
\rho(r) =  \frac{r}{\beta_2} \left(\frac{r}{r_g}\left(1+  \sqrt{1-\frac{r_g}{r}}\right)+\frac{3}{2}(2\beta_2-1)  \right) +{\cal O}(m^2) \, .
\end{array}
\eea
As expected, in the next-to-minimal model, there are coordinate-invariant singularities at the horizon.  As $r \rightarrow r_g$, the relevant scalars become
\be
\label{Tsing}
\sqrt{g^{-1}f}^{\,\mu}_{~~\mu} \rightarrow \frac{1}{\sqrt{\frac{r}{r_g}-1}} +{\cal O}(m^2) \, , ~~~~
{\cal T}^{\mu}_{~~\mu} \rightarrow \frac{2\,(1+3 \beta_2)}{\sqrt{\frac{r}{r_g}-1}}+{\cal O}(m^2)  \, .
\ee
If we consider the full equations of motion \eqref{EOM} away from the massless limit, the singularity in ${\cal T}^{\mu}_{~~\mu} $ means that the curvature $R$ will be singular at the horizon for arbitrarily small $m$ (assuming $\beta_2 \neq -\tfrac{1}{3}$).  It's possible that physical black holes in massive gravity may indeed have coordinate invariant singularities at the horizon.  However, this result indicates that our small $m$ expansion is faulty since the singularity appears at nonzero but arbitrarily small $m$.

\subsubsection{Non-Minimal Model: $\beta_2 \neq 0$, $\beta_3 \neq 0$}
For completeness, we consider also the non-minimal model, when $\beta_2 \neq 0$, $\beta_3 \neq 0$.  To solve the equations of motion for arbitrary $\beta_n$ we take the third equation in \eqref{EOM2} and write it as a quadratic equation for $\rho(r)$:
\bea
\label{nonminrho}
\begin{array}{rcl}
\beta_3 B'_0(r)\rho(r)^2+2 \big( \beta_2 \big(r B'_0(r)+B_0(r)(1-B_1(r))\big)+ \beta_3(1-B_1(r)) \big) \rho(r)&& \\
+\, 2 \beta_2r(1-B_1(r)) +  \beta_1r \big(r B'_0(r)+2B_0(r)(1-B_1(r))\big)&=&0 \, .
\end{array}
\eea
Then, in the massless limit, we have a solution of the form
\bea
\label{EOMnonmin}
\begin{array}{l}
B_0(r) =  \sqrt{1-\frac{r_g}{r}}+{\cal O}(m^2) \, , \vspace{0.3cm} \\
B_1(r)  = \frac{1}{\sqrt{1-\frac{r_g}{r}}} +{\cal O}(m^2)  \, ,
\end{array}
\eea
with $\rho(r)$ given by one of the two roots of equation \eqref{nonminrho}.  We see that the non-minimal model also (naively) supports a Vainshtein mechanism in the strictly massless limit.

However, again there are coordinate invariant singularities at the horizon.  As $r \rightarrow r_g$, the relevant scalars become
\bea
\begin{array}{ccl}
\label{Tnonmin}
\sqrt{g^{-1}f}^{\,\mu}_{~~\mu} & \!\! \rightarrow \!\!&  \frac{1}{\sqrt{\frac{r}{r_g}-1}}  +{\cal O}(m^2) \, ,\vspace{0.2cm} \\
{\cal T}^{\mu}_{~
~\mu} &  \!\! \rightarrow \!\! & \frac{4}{\beta_3} \!\!
\left(\beta_3-\beta_2^2  + \beta_2\beta_3-\beta_3^2 \pm \frac{2\beta_2^3 - 2 \beta_2 \beta_3 + 
 3 \beta_2^2 \beta_3 + \beta_3^2 + 
 12 \beta_2 \beta_3^2 - \beta_3^3}{2 \sqrt{\beta_2^2 + 
  2 \beta_2 \beta_3 -\beta_3  + 5 \beta_3^3}} \right)
 \frac{1}{\sqrt{\frac{r}{r_g}-1}}  +{\cal O}(m^2)\, .
\end{array}
\eea
The two values for ${\cal T}^{\mu}_{~~\mu}$ correspond to the two roots of equation \eqref{nonminrho}.  In the limit that $\beta_3 \rightarrow 0$, this result recovers the the next-to-minimal model result \eqref{Tsing} with the $+$ solution corresponding to $\beta_2>0$ and the $-$ solution corresponding to $\beta_2<0$.  Once again, these results imply that the curvature $R$ generically becomes singular at the horizon at finite $m$.

\section{Time-Dependent Solutions}
\label{timedep}
\subsection{The Massless Limit}
We wish to determine if it is possible to have black hole solutions in massive gravity which have no singularities at the horizon and can still exhibit the expected Yukawa behavior at large distances.  To search for such solutions, we must generalize our ansatz \eqref{ansatz} and relax the requirement of time-independence.  In particular, we will adopt an ansatz where, in the massless limit, the dynamical metric is exactly Schwarzschild and the reference metric is a {\it time-dependent} coordinate transformation of Minkowski.  Focusing on the next-to-minimal model, we demonstrate that this ansatz indeed solves the massive gravity equations of motion and that the singularity at the horizon appears to be avoided.

The coordinates $\tau$ and $\rho$ are the coordinates in which the reference metric is Minkowski.  We now conjecture that there is a different set of coordinates $t$ and $r$ in which the dynamical metric is Schwarzschild in the massless limit.  The original coordinates $\tau$ and $\rho$ are then functions of $t$ and $r$: $\tau = \tau(t,r)$ and $\rho = \rho(t,r)$.  In the $t$ and $r$ coordinates, in the massless limit, the two metrics take the form:
\bea
\begin{array}{lcl}
ds_g^2 &=& -\left(1-\frac{r_g}{r}\right) dt^2+ \frac{1}{1-\frac{r_g}{r}}dr^2+ r^2 \, d\Omega^2  \, , \vspace{0.3cm}\\
ds_f^2 &=& - [\dot{\tau}(t,r)^2-\dot{\rho}(t,r)^2] dt^2+2[\dot{\rho}(t,r)\rho'(t,r)-\dot{\tau}(t,r)\tau'(t,r)]dt dr\\
&&+[\rho'(t,r)^2-\tau'(t,r)^2]dr^2+ \rho(t,r)^2\, d\Omega^2  \, .
\end{array}
\eea
Here, dots denote derivatives with respect to $t$ and primes denote derivatives with respect to $r$.  

We take this ansatz and plug it into the equations of motion \eqref{EOM}.  For simplicity, we consider the next-to-minimal model so that $\beta_3 = 0$.  Taking the massless limit, we find that this is indeed a solution.  The two functions $\tau(t,r)$ and $\rho(t,r)$ are determined by two mixed second-order partial differential equations.  Here we present perturbative solutions for these two functions.    Also for simplicity, we will assume that that $\beta_1>0$ and $\beta_2>0$.   The solutions then naturally organize into a perturbative expansion of the following form:
\bea
\label{soloutin}
\begin{array}{ll}
{\rm For}~ r>r_g: & \tau(t,r)  = \sqrt{3}\, r_g \frac{\beta_1}{\beta_2} \sum_{n=1}^{\infty} \, x\, \tau_n^+ (x) \cosh^n \!\!\left[\frac{t}{2r_g}\right] \left(\frac{r}{r_g}-1\right)^{\!n/2} \, , \\
& \rho(t,r)  =  \sqrt{3}\, r_g \frac{\beta_1}{\beta_2} \sum_{n=1}^{\infty} \, \rho_n^+ (x)  \cosh^n \!\!\left[\frac{t}{2r_g}\right]  \left(\frac{r}{r_g}-1\right)^{\!n/2} \, ,\\
\\
{\rm For}~ r<r_g: &\tau(t,r)  =  \sqrt{3}\, r_g \frac{\beta_1}{\beta_2} \sum_{n=1}^{\infty} \, \tau_n^- (x) \cosh^n \!\!\left[\frac{t}{2r_g}\right]  \left(1-\frac{r}{r_g}\right)^{\!n/2} \, , \\
& \rho(t,r)  =  \sqrt{3}\, r_g \frac{\beta_1}{\beta_2} \sum_{n=1}^{\infty} \, x\, \rho_n^- (x) \cosh^n \!\!\left[\frac{t}{2r_g}\right]\left(1-\frac{r}{r_g}\right)^{\!n/2} \, ,
\end{array}
\eea
where we have defined $x \equiv \tanh[t/2r_g]$.  To order $n=5$, we find the following solutions for $\tau_n^\pm (x)$ and $\rho_n^\pm (x)$:
\bea
\label{taurho}
\begin{array}{lll}
\tau_1^+(x) = F_1 \, , & ~~~~ & \tau_1^-(x) = F_1 \, ,  \\
\rho_1^+(x) = G_1 \, , & ~~~~ & \rho_1^-(x) = G_1 \, ,\\
\\
\tau_2^+(x) = F_2 \, , & ~~~~ & \tau_2^-(x) = F_2\, x \, ,  \\
\rho_2^+(x) = G_{2a} + G_{2b}\, x^2 \, , & ~~~~ & \rho_2^-(x) = G_{2a}\, x + G_{2b}\, x^{-1} \, , \\
\\
\tau_3^+(x) = F_{3a} + F_{3b} \, x^2 \, , & ~~~~ & \tau_3^-(x) = F_{3a}\, x^2+F_{3b} \, ,  \\
\rho_3^+(x) = G_{3a} + G_{3b}\, x^2 \, , & ~~~~ & \rho_3^-(x) = G_{3a}\, x^2 + G_{3b}\, ,  \\
\\
\tau_4^+(x) = F_{4a} + F_{4b} \, x^2 \, , & ~~~~ & \tau_4^-(x) = F_{4a}\, x^3+F_{4b}\, x \, ,  \\
\rho_4^+(x) = G_{4a} + G_{4b}\, x^2 + G_{4c}\, x^4 \, , & ~~~~ & \rho_4^-(x) = G_{4a}\, x^3 + G_{4b}\, x+ G_{4c}\, x^{-1} \, , \\
\\
\tau_5^+(x) = F_{5a} + F_{5b} \, x^2+ F_{5c} \, x^4 \, , & ~~~~ & \tau_5^-(x) = F_{5a}\, x^4+F_{5b}\, x^2 +F_{5c}\, ,  \\
\rho_5^+(x) = G_{5a} + G_{5b}\, x^2 + G_{5c}\, x^4 \, , & ~~~~ & \rho_5^-(x) = G_{5a}\, x^4 + G_{5b}\, x^2+ G_{5c} \, , \\
\end{array}
\eea
where
\bea
\label{FG}
\begin{array}{lclclcl}
F_1 &\!\!=\!\!& 1 \, , & ~~~~~~ & F_{4a} &\!\!=\!\!& - \,8 \,\frac{20212216 + 10761615 \sqrt{3}}{3987555} \, ,\\
G_1 &\!\!=\!\!& 1\, , & ~~~~~~ & F_{4b} &\!\!=\!\!& - \,4\, \frac{86474551 + 52074576 \sqrt{3}}{3987555} \, ,\\
&&                  & ~~~~~~ & G_{4a} &\!\!=\!\!& -\,\frac{13859147 - 13307623 \sqrt{3}}{7975110} \, ,\\
&&                  & ~~~~~~ & G_{4b} &\!\!=\!\!&-\,7\, \frac{18003911 + 11074631 \sqrt{3}}{1329185} \, ,\\
F_2 &\!\!=\!\!& -\,\frac{2}{13} (19 + 5 \sqrt{3}) \, , & ~~~~ & G_{4c} &\!\!=\!\!& -\,\frac{245168455 + 136955569 \sqrt{3}}{7975110}\, ,\\
G_{2a} &\!\!=\!\!&-\,\frac{1}{13} (5 - 11 \sqrt{3}) \, , \\
G_{2b} &\!\!=\!\!&-\,\frac{3}{13} (11 + 7 \sqrt{3}) \, ,\\
&&                           & ~~~~~~ & F_{5a} &\!\!=\!\!&\frac{2358595986147 + 1341907450280 \sqrt{3}}{14514700200}\, ,\\
&&                           & ~~~~~~ & F_{5b} &\!\!=\!\!&\frac{6207302851481 + 3597373424072 \sqrt{3}}{7257350100}\, ,\\
F_{3a} &\!\!=\!\!&\frac{17367 + 7862 \sqrt{3}}{1690} \, ,  & ~~~~~~ & F_{5c} &\!\!=\!\!& \frac{791752137769 + 452552246456 \sqrt{3}}{4838233400}\, ,\\
F_{3b} &\!\!=\!\!&\frac{27883 + 18662 \sqrt{3}}{5070} \, , & ~~~~~~ & G_{5a} &\!\!=\!\!& \frac{58623861651 - 29657041376 \sqrt{3}}{14514700200}\, ,\\
G_{3a} &\!\!=\!\!&\frac{14305 - 3586 \sqrt{3}}{5070} \, , & ~~~~~~ & G_{5b} &\!\!=\!\!&\frac{32883045033 + 19373615360 \sqrt{3}}{59978100}\, ,\\
G_{3b} &\!\!=\!\!& \frac{21893 + 15278 \sqrt{3}}{1690} \, , & ~~~~~~ & G_{5c} &\!\!=\!\!&\frac{830194303889 + 475959378368 \sqrt{3}}{1319518200}\, .\\
\end{array}
\eea

\begin{figure}[t!]
\centering
\includegraphics[scale=0.3]{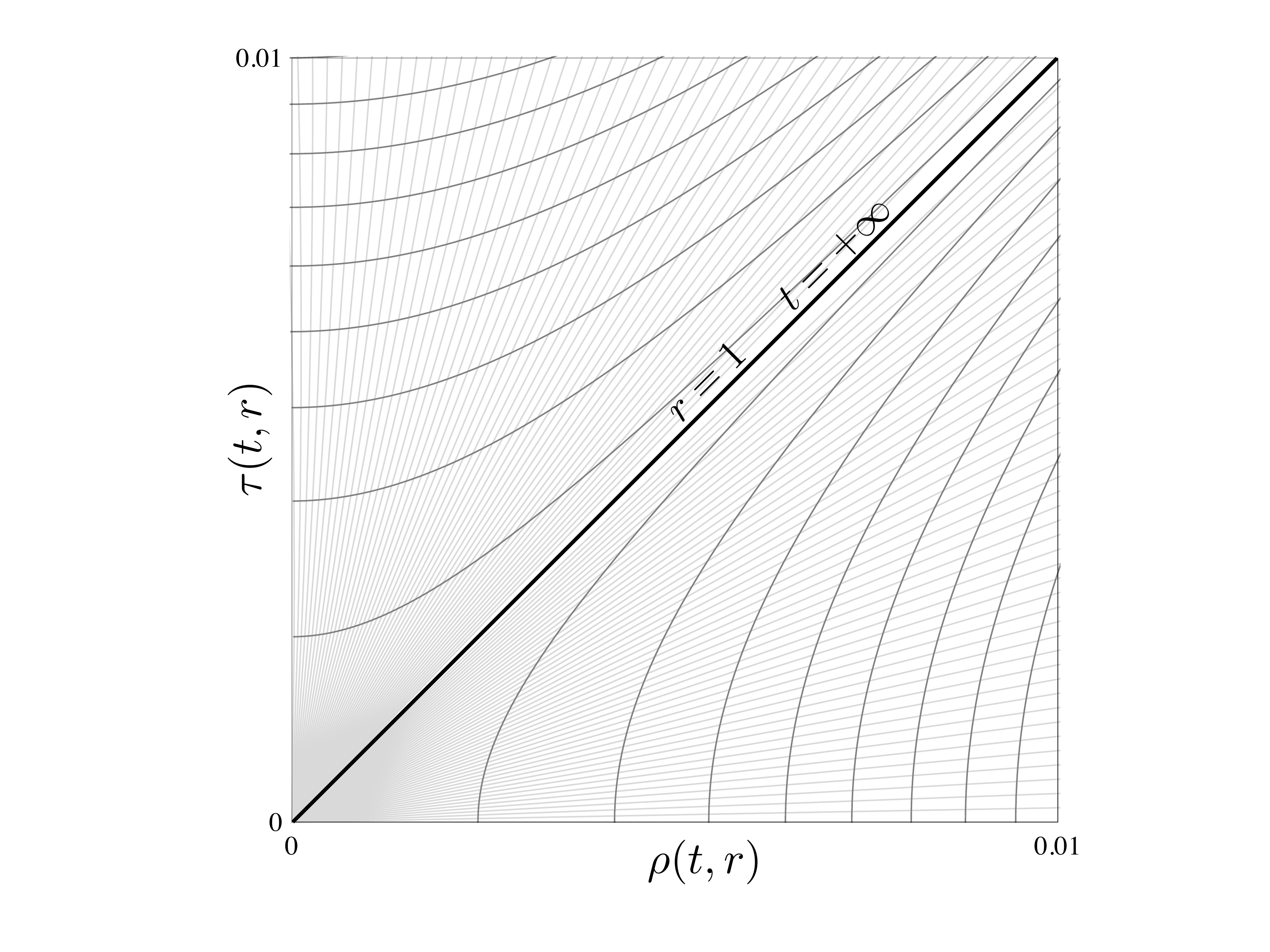}
\caption{$\tau(t,r)$ and $\rho(t,r)$ in units of $r_g$. Lines of constant $r$ are dark gray and lines of constant $t$ are light gray.  The horizon is at $\tau (t,r) = \rho(t,r) $.  $\beta_1/\beta_2$ is set to 1.    }
\end{figure}

When solving, we have imposed two boundary conditions.  First, we have imposed that sending $t \rightarrow - t$ takes $\tau \rightarrow -\tau$ and $\rho \rightarrow \rho$.   Second, we have imposed that $\tau (t,r) = \rho(t,r) $  at the horizon, i.e., when $r =r_g$ and $\tanh[t/2r_g] = 1$.  In other words, we are enforcing that the horizon be a null surface in the $(\tau,\rho)$ coordinates of the reference metric.  These two conditions fix all integration constants at each order in $n$.

We plot the solution in Figure 2.  The dark gray lines are lines of constant $r$ and the light gray lines are lines of constant $t$.  The horizon corresponds to $\tau = \rho$.  Outside the horizon, $t=0$ corresponds to $\tau = 0$.  However, we note that inside the horizon, $t=0$ does {\it not} correspond to $\rho =0$.  Otherwise, these solutions share many similarities with Kruskal-Szekeres coordinates.  In particular, at lowest order in $n$, they are given by
\bea
\begin{array}{ll}
{\rm For}~ r>r_g: & \tau(t,r)  \simeq r_g \frac{\beta_1}{\beta_2} \sqrt{3} \, \left(\frac{r}{r_g}-1\right)^{\! 1/2} \, \sinh\!\left[\frac{t}{2 r_g}\right] \, , \\
& \rho(t,r)  \simeq r_g \frac{\beta_1}{\beta_2} \sqrt{3} \, \left(\frac{r}{r_g}-1\right)^{\! 1/2} \, \cosh\!\left[\frac{t}{2 r_g}\right] \, , \\
\\
{\rm For}~ r<r_g: &\tau(t,r)  \simeq r_g \frac{\beta_1}{\beta_2}\, \sqrt{3} \left(1-\frac{r}{r_g}\right)^{\! 1/2} \, \cosh\!\left[\frac{t}{2 r_g}\right] \, , \\
& \rho(t,r)  \simeq r_g \frac{\beta_1}{\beta_2} \sqrt{3} \, \left(1-\frac{r}{r_g}\right)^{\! 1/2} \,\sinh\!\left[\frac{t}{2 r_g}\right]  \, .
\end{array}
\eea

\subsection{Finiteness at the Horizon}
With this solution, we can check for the presence of coordinate-invariant singularities at the horizon.  In particular, we wish to verify that ${\cal T}^\mu_{~~\mu}$ as defined in \eqref{EOM} is finite at the horizon, indicating that there are no curvature singularities at the horizon at finite $m$.  To do so, we must use variables that are well-defined at the horizon.  Thus, we switch from Schwarzschild coordinates $(t,r)$ to Kruskal-Szekeres coordinates $(T,R)$:
\bea
\begin{array}{ll}
{\rm For}~ r>r_g: & T(t,r)  = {\rm e}^{r/2r_g} \left(\frac{r}{r_g}-1\right)^{\!1/2}  \sinh\!\left[\frac{t}{2r_g}\right]    \, , \\
& R(t,r)  = {\rm e}^{r/2r_g} \left(\frac{r}{r_g}-1\right)^{\!1/2}  \cosh\!\left[\frac{t}{2 r_g}\right]  \, ,\\
\\
{\rm For}~ r<r_g: & T(t,r)  = {\rm e}^{r/2r_g} \left(1-\frac{r}{r_g}\right)^{\!1/2}  \cosh\!\left[\frac{t}{2 r_g}\right]   \, , \\
& R(t,r)  = {\rm e}^{r/2r_g} \left(1-\frac{r}{r_g}\right)^{\!1/2}  \sinh\!\left[\frac{t}{2 r_g}\right]  \, .
\end{array}
\eea

In these coordinates, the expansion parameter of our solutions \eqref{soloutin} becomes
\be
\label{param}
\cosh^n \!\!\left[\frac{t}{2r_g}\right] \left(\frac{r}{r_g}-1\right)^{\! n/2} = \left(\frac{R}{\sqrt{R^2-T^2}}\right)^{\!n}\, W\!\left(\frac{R^2-T^2}{{\rm e}}\right)^{\! n/2} \, ,
\ee
where $W(x)$ is the Lambert-W function.  At the horizon, i.e., in the limit that $R \rightarrow T$, the expansion parameter \eqref{param} becomes $T^n/{\rm e}^{n/2}$.

We calculate ${\cal T}^\mu_{~~\mu}$ at each order and evaluate at the horizon $R =T$.  Up to fifth order for which we have solved, we find that ${\cal T}^\mu_{~~\mu}$ is indeed finite, in contrast to the static results \eqref{Tsing}, \eqref{Tnonmin},
\bea
\begin{array}{lcl}
{\cal T}^\mu_{~~\mu}&\!\! \rightarrow \!\! & 4 \beta_0+ \frac{3}{2}(1+2\sqrt{3})\frac{\beta_1^2}{\beta_2} 
 +\frac{6}{13} (29 + 22 \sqrt{3}) \frac{\beta_1^2}{\beta_2}  \frac{T}{{\rm e}^{1/2}}  \vspace{0.2cm} \\
&&+ \frac{18}{845} (2129 + 251 \sqrt{3})  \frac{\beta_1^2}{\beta_2}  \frac{T^2}{{\rm e}} 
+4\, \frac{379366904 + 215110657 \sqrt{3}}{1329185}\frac{\beta_1^2}{\beta_2} \frac{T^3}{{\rm e}^{3/2}}  \vspace{0.2cm}\\
&& +2\, \frac{2084180619478 + 1202578322999 \sqrt{3}}{604779175} \frac{\beta_1^2}{\beta_2}  \frac{T^4}{{\rm e^2}} 
+ {\cal O}\! \left(\frac{T^5}{{\rm e}^{5/2}} \right) \, .  \\
\end{array}
\eea
If the pattern of functions given in \eqref{taurho} continues as expected, ${\cal T}^\mu_{~~\mu}$ will remain finite at each order in $T^n/{\rm e}^{n/2}$.  Interestingly, the value of ${\cal T}^\mu_{~~\mu}$ at the horizon depends explicitly on the Kruskal-Szekeres time $T$.  We will see below that this doesn't necessarily indicated that the apparent horizon of the black hole is changing as a function of $T$.

To summarize, we see that a time-dependent ansatz allows for black hole solutions in massive gravity which smoothly approach Schwarzschild black holes in the massless limit and which do not appear to exhibit coordinate-invariant singularities at the horizon.

\subsection{Nonzero Mass and Apparent Horizon}
Let us now consider the leading order $m^2$ corrections to the dynamical metric.  Since the quantity ${\cal T}_{\mu\nu}$ depends explicitly on the Schwarzschild time coordinate $t$, from the equations of motion \eqref{EOM}, we anticipate that the $m^2$ corrections to the dynamical metric will be time-dependent as well.

A generic time-dependent, spherically symmetric metric can be put in the form
\be
\label{gtime}
ds^2 = -e^{2 \Phi(t,r)} \left(\text{\footnotesize$1-\frac{2 G M(t,r)}{r} $}\right) dt^2 + \frac{1}{1-\frac{2 G M(t,r)}{r}}dr^2 +r^2 d\Omega^2 \, ,
\ee
defined by two functions, $\Phi(t,r)$ and $M(t,r)$.  The function $M(t,r)$ is the Misner-Sharp mass function which corresponds to the quasi-local mass contained within a sphere of radius $r$ at time $t$.  We take expression \eqref{gtime} to be our ansatz for the dynamical metric $g_{\mu\nu}$ at finite $m^2$.  Using the equations of motion \eqref{EOM} along with our solutions \eqref{soloutin}-\eqref{FG}, we can determine $\Phi(t,r)$ and $M(t,r)$ to leading order in $m^2$.  For the physically interesting quantity $M(t,r)$, we find, to leading order in $m^2$ and for $r>r_g$:
\be
\label{finitem}
\frac{2\, G M(t,r)}{r_g} \simeq 1+ m^2 r_g^2 \frac{\beta_1^2}{\beta_2}\left[
\sum_{n=3}^\infty c_n (x) \cosh^n \!\!\left[\frac{t}{2r_g}\right] \left(\frac{r}{r_g}-1\right)^{n/2}
+\sum_{n=1}^\infty D_n\left( \frac{r}{r_g}-1 \right)^n  \right] \, ,
\ee
where again $x \equiv \tanh[t/2r_g] $.  The first few terms are determined to be
\bea
\label{c}
\begin{array}{ccl}
c_3(x) &=& C_{3a} + C_{3b} \, x^2 \, , \\
c_4(x) &=& C_{4a} + C_{4b}\, x^2 \, ,\\
c_5(x) &=& C_{5a} + C_{5b} \,x^2+ C_{5c}\, x^4 \, , \\
c_6(x) &=& C_{6a} + C_{6b}\, x^2 +C_{6c}\, x^4 \, ,\\
\end{array}
\eea
with constants given by
\bea
\label{C}
\begin{array}{cclcccl}
C_{3a} &=& \frac{2}{13} (24 + 7 \sqrt{3})\, , & ~~~~~~ &C_{4a} &=&-\,\frac{2}{845} (4632 + 727 \sqrt{3})\, , \\
C_{3b} &=& -\, \frac{2}{13} (24 + 7 \sqrt{3}) \, , & ~~~~~~ &C_{4b} &=& \frac{2}{845} (4632 + 727 \sqrt{3})\, , \\
\\
C_{5a} &=& \frac{89961180 - 5942257 \sqrt{3}}{3987555}\, , & ~~~~~~ &
C_{6a} &=&\frac{1980193455 + 13177986034 \sqrt{3}}{362867505}\, , \\
C_{5b} &=& 2\, \frac{751378 + 13863771 \sqrt{3}}{1329185} \, , & ~~~~~~ &
C_{6b} &=&-\,\frac{7679972382 + 5981929537 \sqrt{3}}{27912885}\, , \\
C_{5c} &=& -\, \frac{94469448 + 77240369 \sqrt{3}}{3987555} \, , & ~~~~~~ &
C_{6c} &=& \frac{32619815837 + 21529032649 \sqrt{3}}{120955835}\, , \\
\end{array}
\eea
and
\bea
\begin{array}{ccl}
D_1 &=& \frac{\beta_0\beta_2}{\beta_1^2} +\frac{\sqrt{3}}{2} \, , \vspace{0.2cm} \\
D_2 &=& \frac{\beta_0\beta_2}{\beta_1^2} + 3\, \frac{13159 + 8602 \sqrt{3}}{3380}\, , \vspace{0.2cm}\\
D_3 &=& \frac{\beta_0\beta_2}{3\beta_1^2} - \frac{99529913559 + 49389993413 \sqrt{3}}{4354410060}\, .\\
\end{array}
\eea
The apparent horizon of the time-dependent black hole is given by the implicit condition $2 G M(t,r_H) = r_H$.  However, in Schwarzschild coordinates, this condition is ill-defined because $t$ is not a good coordinate at the horizon.  Thus, to determine the time dependence of the horizon, it is necessary to switch to better behaved coordinates.   

This can be done generically by adopting an Eddington-Finkelstein type of time coordinate $t \rightarrow v(t,r)$ so that the metric \eqref{gtime} becomes
\be
ds^2 = -F(v,r)^2 \left(1-\frac{2 G M(v,r)}{r} \right) dv^2 +2 F(v,r) dv dr +r^2 d\Omega^2 \, .
\ee
The coordinate $v$ is well-behaved at the horizon.  Thus the condition $2 G M(v,r_H) = r_H$ can be used to determine the position of the apparent horizon as a function of $v$: $r_H(v)$.  However, given the nature of our solutions \eqref{finitem}, finding $M(v,r)$ is nontrivial so we will take a simpler approach.

Let us assume that the position of the apparent horizon is, in fact, time-{\it independent}.  In other words, we assume that $r_H(\tilde{t}) = r_g$ for some coordinate $\tilde{t}$ that is regular at the horizon.  To verify this, we should adopt a time coordinate that is well-behaved at $r=r_g$.  In particular, we can adopt the Kruskal-Szekeres time coordinate $T$: $t \rightarrow T(t,r)$.  Then, the Misner-Sharp mass defined in \eqref{finitem} and evaluated at the horizon $r=r_g$ becomes
\be
\frac{2\, G M(T,r=r_g)}{r_g} = 1+ m^2 r_g^2 \frac{\beta_1^2}{\beta_2}\sum_{n=3}^\infty c_n (x=1) \frac{T^n}{{\rm e}^{n/2}} + {\cal O}(m^4) \, .
\ee
We see that our assumption is valid as long as $c_n(x=1) = 0$ at each order in $n$.  Comparing with our expressions \eqref{c} and \eqref{C} it is easily verified that this is indeed the case and that and $r_H(T) = r_g$.  The position of the apparent horizon is time-{\it independent} at leading order in $m^2$.  The implication is that, for this solution, the black hole is neither accreting nor evaporating, despite the solution being explicitly time-dependent.

\section{Discussion}
We have demonstrated that time-dependent black hole solutions in massive gravity can potentially evade the problem of coordinate-invariant singularities at the horizon and can smoothly recover the black hole solutions of General Relativity in the massless limit.  The solutions that we have found are derived in the limit of small graviton mass and are thus valid only well inside the Vainshtein radius.  Thus, it remains to be seen if these solutions can be matched to the expected Yukawa asymptotics at large $r$.  Such a matching is certainly allowed by the given parameter space.  I.e., the Fierz-Pauli mass is {\it not} set to zero on these solutions.  It might appear unrealistic to expect that time-dependent solutions could be matched to static asymptotics.  However, given that the apparent horizon of these solutions is, in fact, static at leading order in $m^2$, it is physically intuitive that the asymptotics might reflect a black hole that is neither accreting or evaporating.

Finally, we note that the solutions we have derived are not necessarily unique.  It is possible that other time-dependent solutions exists, perhaps with features even more desirable than those found here.  Ultimately, however, a physical black hole formed from gravitational collapse will be described by a particular branch of solutions.  While the time-dependent solutions found here resolve the potentially problematic properties of the static solutions, it remains undetermined what is the correct physical branch.

\vskip.5cm

\bigskip
{\bf Acknowledgements}: 
I would like to thank Cedric Deffayet, Kurt Hinterbichler, Lam Hui, Austin Joyce and Riccardo Penco for many fruitful discussions. This work was supported by DOE grant DE-SC0011941 and NASA grant NNX16AB27G.

\bibliographystyle{utphys}
\addcontentsline{toc}{section}{References}
\bibliography{blackholes}

\providecommand{\href}[2]{#2}\begingroup\raggedright\begin{thebibliography}{10}

\bibitem{deRham:2010ik}
C.~de~Rham and G.~Gabadadze, ``{Generalization of the Fierz-Pauli Action},''
  \href{http://dx.doi.org/10.1103/PhysRevD.82.044020}{{\em Phys. Rev.}
  {\bfseries D82} (2010) 044020},
\href{http://arxiv.org/abs/1007.0443}{{\ttfamily arXiv:1007.0443 [hep-th]}}.

\bibitem{deRham:2010kj}
C.~de~Rham, G.~Gabadadze, and A.~J. Tolley, ``{Resummation of Massive
  Gravity},'' \href{http://dx.doi.org/10.1103/PhysRevLett.106.231101}{{\em
  Phys.Rev.Lett.} {\bfseries 106} (2011) 231101},
\href{http://arxiv.org/abs/1011.1232}{{\ttfamily arXiv:1011.1232 [hep-th]}}.

\bibitem{Boulware:1973my}
D.~G. Boulware and S.~Deser, ``{Can gravitation have a finite range?},''
\href{http://dx.doi.org/10.1103/PhysRevD.6.3368}{{\em Phys. Rev.} {\bfseries
  D6} (1972) 3368--3382}.

\bibitem{Hassan:2011hr}
S.~Hassan and R.~A. Rosen, ``{Resolving the Ghost Problem in non-Linear Massive
  Gravity},'' \href{http://dx.doi.org/10.1103/PhysRevLett.108.041101}{{\em
  Phys.Rev.Lett.} {\bfseries 108} (2012) 041101},
\href{http://arxiv.org/abs/1106.3344}{{\ttfamily arXiv:1106.3344 [hep-th]}}.

\bibitem{Hinterbichler:2011tt}
K.~Hinterbichler, ``{Theoretical Aspects of Massive Gravity},''
\href{http://arxiv.org/abs/1105.3735}{{\ttfamily arXiv:1105.3735 [hep-th]}}.

\bibitem{deRham:2014zqa}
C.~de~Rham, ``{Massive Gravity},''
  \href{http://dx.doi.org/10.12942/lrr-2014-7}{{\em Living Rev.Rel.} {\bfseries
  17} (2014) 7},
\href{http://arxiv.org/abs/1401.4173}{{\ttfamily arXiv:1401.4173 [hep-th]}}.

\bibitem{Koyama:2011xz}
K.~Koyama, G.~Niz, and G.~Tasinato, ``{Analytic solutions in non-linear massive
  gravity},'' \href{http://dx.doi.org/10.1103/PhysRevLett.107.131101}{{\em
  Phys. Rev. Lett.} {\bfseries 107} (2011) 131101},
\href{http://arxiv.org/abs/1103.4708}{{\ttfamily arXiv:1103.4708 [hep-th]}}.

\bibitem{Nieuwenhuizen:2011sq}
T.~M. Nieuwenhuizen, ``{Exact Schwarzschild-de Sitter black holes in a family
  of massive gravity models},''
  \href{http://dx.doi.org/10.1103/PhysRevD.84.024038}{{\em Phys. Rev.}
  {\bfseries D84} (2011) 024038},
\href{http://arxiv.org/abs/1103.5912}{{\ttfamily arXiv:1103.5912 [gr-qc]}}.

\bibitem{Koyama:2011yg}
K.~Koyama, G.~Niz, and G.~Tasinato, ``{Strong interactions and exact solutions
  in non-linear massive gravity},''
  \href{http://dx.doi.org/10.1103/PhysRevD.84.064033}{{\em Phys. Rev.}
  {\bfseries D84} (2011) 064033},
\href{http://arxiv.org/abs/1104.2143}{{\ttfamily arXiv:1104.2143 [hep-th]}}.

\bibitem{Gruzinov:2011mm}
A.~Gruzinov and M.~Mirbabayi, ``{Stars and Black Holes in Massive Gravity},''
  \href{http://dx.doi.org/10.1103/PhysRevD.84.124019}{{\em Phys. Rev.}
  {\bfseries D84} (2011) 124019},
\href{http://arxiv.org/abs/1106.2551}{{\ttfamily arXiv:1106.2551 [hep-th]}}.

\bibitem{Comelli:2011wq}
D.~Comelli, M.~Crisostomi, F.~Nesti, and L.~Pilo, ``{Spherically Symmetric
  Solutions in Ghost-Free Massive Gravity},''
  \href{http://dx.doi.org/10.1103/PhysRevD.85.024044}{{\em Phys. Rev.}
  {\bfseries D85} (2012) 024044},
\href{http://arxiv.org/abs/1110.4967}{{\ttfamily arXiv:1110.4967 [hep-th]}}.

\bibitem{Berezhiani:2011mt}
L.~Berezhiani, G.~Chkareuli, C.~de~Rham, G.~Gabadadze, and A.~J. Tolley, ``{On
  Black Holes in Massive Gravity},''
  \href{http://dx.doi.org/10.1103/PhysRevD.85.044024}{{\em Phys. Rev.}
  {\bfseries D85} (2012) 044024},
\href{http://arxiv.org/abs/1111.3613}{{\ttfamily arXiv:1111.3613 [hep-th]}}.

\bibitem{Volkov:2012wp}
M.~S. Volkov, ``{Hairy black holes in the ghost-free bigravity theory},''
  \href{http://dx.doi.org/10.1103/PhysRevD.85.124043}{{\em Phys. Rev.}
  {\bfseries D85} (2012) 124043},
\href{http://arxiv.org/abs/1202.6682}{{\ttfamily arXiv:1202.6682 [hep-th]}}.

\bibitem{Sbisa:2012zk}
F.~Sbisa, G.~Niz, K.~Koyama, and G.~Tasinato, ``{Characterising Vainshtein
  Solutions in Massive Gravity},''
  \href{http://dx.doi.org/10.1103/PhysRevD.86.024033}{{\em Phys. Rev.}
  {\bfseries D86} (2012) 024033},
\href{http://arxiv.org/abs/1204.1193}{{\ttfamily arXiv:1204.1193 [hep-th]}}.

\bibitem{Gratia:2012wt}
P.~Gratia, W.~Hu, and M.~Wyman, ``{Self-accelerating Massive Gravity: Exact
  solutions for any isotropic matter distribution},''
  \href{http://dx.doi.org/10.1103/PhysRevD.86.061504}{{\em Phys. Rev.}
  {\bfseries D86} (2012) 061504},
\href{http://arxiv.org/abs/1205.4241}{{\ttfamily arXiv:1205.4241 [hep-th]}}.

\bibitem{Chiang:2012vh}
C.-I. Chiang, K.~Izumi, and P.~Chen, ``{Spherically symmetric analysis on open
  FLRW solution in non-linear massive gravity},''
  \href{http://dx.doi.org/10.1088/1475-7516/2012/12/025}{{\em JCAP} {\bfseries
  1212} (2012) 025},
\href{http://arxiv.org/abs/1208.1222}{{\ttfamily arXiv:1208.1222 [hep-th]}}.

\bibitem{Mirbabayi:2013sva}
M.~Mirbabayi and A.~Gruzinov, ``{Black hole discharge in massive
  electrodynamics and black hole disappearance in massive gravity},''
  \href{http://dx.doi.org/10.1103/PhysRevD.88.064008}{{\em Phys. Rev.}
  {\bfseries D88} (2013) 064008},
\href{http://arxiv.org/abs/1303.2665}{{\ttfamily arXiv:1303.2665 [hep-th]}}.

\bibitem{Volkov:2013roa}
M.~S. Volkov, ``{Self-accelerating cosmologies and hairy black holes in
  ghost-free bigravity and massive gravity},''
  \href{http://dx.doi.org/10.1088/0264-9381/30/18/184009}{{\em Class. Quant.
  Grav.} {\bfseries 30} (2013) 184009},
\href{http://arxiv.org/abs/1304.0238}{{\ttfamily arXiv:1304.0238 [hep-th]}}.

\bibitem{Tasinato:2013rza}
G.~Tasinato, K.~Koyama, and G.~Niz, ``{Exact Solutions in Massive Gravity},''
  \href{http://dx.doi.org/10.1088/0264-9381/30/18/184002}{{\em Class. Quant.
  Grav.} {\bfseries 30} (2013) 184002},
\href{http://arxiv.org/abs/1304.0601}{{\ttfamily arXiv:1304.0601 [hep-th]}}.

\bibitem{Babichev:2013una}
E.~Babichev and A.~Fabbri, ``{Instability of black holes in massive gravity},''
  \href{http://dx.doi.org/10.1088/0264-9381/30/15/152001}{{\em Class. Quant.
  Grav.} {\bfseries 30} (2013) 152001},
\href{http://arxiv.org/abs/1304.5992}{{\ttfamily arXiv:1304.5992 [gr-qc]}}.

\bibitem{Brito:2013xaa}
R.~Brito, V.~Cardoso, and P.~Pani, ``{Black holes with massive graviton
  hair},'' \href{http://dx.doi.org/10.1103/PhysRevD.88.064006}{{\em Phys. Rev.}
  {\bfseries D88} (2013) 064006},
\href{http://arxiv.org/abs/1309.0818}{{\ttfamily arXiv:1309.0818 [gr-qc]}}.

\bibitem{Arraut:2013bqa}
I.~Arraut, ``{On the black holes in alternative theories of gravity: The case
  of nonlinear massive gravity},''
  \href{http://dx.doi.org/10.1142/S0218271815500224}{{\em Int. J. Mod. Phys.}
  {\bfseries D24} (2015) 1550022},
\href{http://arxiv.org/abs/1311.0732}{{\ttfamily arXiv:1311.0732 [gr-qc]}}.

\bibitem{Kodama:2013rea}
H.~Kodama and I.~Arraut, ``{Stability of the Schwarzschild--de Sitter black
  hole in the dRGT massive gravity theory},''
  \href{http://dx.doi.org/10.1093/ptep/ptu016}{{\em PTEP} {\bfseries 2014}
  (2014) 023E02},
\href{http://arxiv.org/abs/1312.0370}{{\ttfamily arXiv:1312.0370 [hep-th]}}.

\bibitem{Renaux-Petel:2014pja}
S.~Renaux-Petel, ``{On the Vainshtein mechanism in the minimal model of massive
  gravity},'' \href{http://dx.doi.org/10.1088/1475-7516/2014/03/043}{{\em JCAP}
  {\bfseries 1403} (2014) 043},
\href{http://arxiv.org/abs/1401.0497}{{\ttfamily arXiv:1401.0497 [hep-th]}}.

\bibitem{Babichev:2014oua}
E.~Babichev and A.~Fabbri, ``{Stability analysis of black holes in massive
  gravity: a unified treatment},''
  \href{http://dx.doi.org/10.1103/PhysRevD.89.081502}{{\em Phys. Rev.}
  {\bfseries D89} no.~8, (2014) 081502},
\href{http://arxiv.org/abs/1401.6871}{{\ttfamily arXiv:1401.6871 [gr-qc]}}.

\bibitem{Volkov:2014ooa}
M.~S. Volkov, ``{Hairy black holes in theories with massive gravitons},''
  \href{http://dx.doi.org/10.1007/978-3-319-10070-8_6}{{\em Lect. Notes Phys.}
  {\bfseries 892} (2015) 161--180},
\href{http://arxiv.org/abs/1405.1742}{{\ttfamily arXiv:1405.1742 [hep-th]}}.

\bibitem{Babichev:2015xha}
E.~Babichev and R.~Brito, ``{Black holes in massive gravity},''
  \href{http://dx.doi.org/10.1088/0264-9381/32/15/154001}{{\em Class. Quant.
  Grav.} {\bfseries 32} (2015) 154001},
\href{http://arxiv.org/abs/1503.07529}{{\ttfamily arXiv:1503.07529 [gr-qc]}}.

\bibitem{Deffayet:2011rh}
C.~Deffayet and T.~Jacobson, ``{On horizon structure of bimetric spacetimes},''
  \href{http://dx.doi.org/10.1088/0264-9381/29/6/065009}{{\em Class. Quant.
  Grav.} {\bfseries 29} (2012) 065009},
\href{http://arxiv.org/abs/1107.4978}{{\ttfamily arXiv:1107.4978 [gr-qc]}}.

\bibitem{Fierz:1939ix}
M.~Fierz and W.~Pauli, ``{On relativistic wave equations for particles of
  arbitrary spin in an electromagnetic field},''
{\em Proc. Roy. Soc. Lond.} {\bfseries A173} (1939) 211--232.

\bibitem{vanDam:1970vg}
H.~van Dam and M.~J.~G. Veltman, ``{Massive and massless Yang-Mills and
  gravitational fields},''
\href{http://dx.doi.org/10.1016/0550-3213(70)90416-5}{{\em Nucl. Phys.}
  {\bfseries B22} (1970) 397--411}.

\bibitem{Zakharov:1970cc}
V.~Zakharov, ``{Linearized gravitation theory and the graviton mass},''
{\em JETP Lett.} {\bfseries 12} (1970) 312.

\bibitem{Vainshtein:1972sx}
A.~I. Vainshtein, ``{To the problem of nonvanishing gravitation mass},''
\href{http://dx.doi.org/10.1016/0370-2693(72)90147-5}{{\em Phys. Lett.}
  {\bfseries B39} (1972) 393--394}.

\bibitem{Hassan:2011vm}
S.~Hassan and R.~A. Rosen, ``{On Non-Linear Actions for Massive Gravity},''
  \href{http://dx.doi.org/10.1007/JHEP07(2011)009}{{\em JHEP} {\bfseries 1107}
  (2011) 009},
\href{http://arxiv.org/abs/1103.6055}{{\ttfamily arXiv:1103.6055 [hep-th]}}.

\end{thebibliography}\endgroup

\end{document}